\documentclass[aps,prl,twocolumn,superscriptaddress]{revtex4}
\usepackage{graphicx}% Include figure files

\begin{document}

%\preprint{}

\title{Neutrino signals from the formation of black hole: \\
A probe of equation of state of dense matter}

% 7/25, intro
% 7/26, model
% 7/27pm, data analysis, comparison of two runs, profiles
% 8/5pm, data analysis, comparison of two runs, trajectories
% 8/8, results
% 8/9, data analysis and results
% 8/10, summary
% 8/12pm, figure and references
% 8/30pm, data, figure updated
% 8/31pm, polish figure, data
% 9/3evening, edit sentences and shorten
% 9/10am, shorten and strengthen
% 9/16pm, revise by Yamada


\author{K. Sumiyoshi}
\email[]{sumi@numazu-ct.ac.jp}
%\thanks{}
\affiliation{Numazu College of Technology, Ooka 3600, Numazu, Shizuoka 410-8501, Japan}
\affiliation{National Astronomical Observatory of Japan, 2-21-1 Osawa, Mitaka, Tokyo 181-8588, Japan}

\author{S. Yamada}
\affiliation{Science \& Engineering and Advanced Research Institute for Science and Engineering, Waseda University,
             Okubo, 3-4-1, Shinjuku, Tokyo 169-8555, Japan}

\author{H. Suzuki}
\affiliation{Faculty of Science and Technology, Tokyo University of Science,
             Yamazaki 2641, Noda, Chiba 278-8510, Japan}

\author{S. Chiba}
\affiliation{Advanced Science Research Center, Japan Atomic Energy Research Institute, 
             Tokai, Ibaraki 319-1195, Japan}


\date{\today}

\begin{abstract}
The gravitational collapse of a non-rotating, black-hole-forming massive star is 
studied by $\nu$-radiation-hydrodynamical simulations for two different sets of realistic 
equation of state of dense matter.  
We show that the event will produce as many neutrinos as the ordinary supernova, but 
with distinctive characteristics in luminosities and spectra that will be an unmistakable 
indication of black hole formation. More importantly, the neutrino signals are quite 
sensitive to the difference of equation of state and can be used as a useful probe
into the properties of dense matter. The event will be unique in that they will be shining 
only by neutrinos (and, possibly, gravitational waves) but not by photons, and hence they 
should be an important target of neutrino astronomy. 
\end{abstract}

%\keywords{}

\maketitle

%\section{}
%\subsection{}
%\subsubsection{}

% If in two-column mode, this environment will change to single-column
% format so that long equations can be displayed. Use
%\begin{widetext}
%\end{widetext}

Massive stars of $\gtrsim$10 solar masses (M$_\odot$) end 
their lives when the iron core is formed at the center \cite{bet90}.  
For the mass range of $\sim$10--20M$_\odot$, the spectacular optical display 
known as the supernova explosion occurs following the gravitational 
collapse of iron core and the launch of shock wave by the 
bounce of core at high densities.  
The successful explosion produces a proto-neutron star 
that emits a bunch of neutrinos for $\sim$20 s during the formation and cool 
down \cite{suz94}, which was vindicated in SN 1987A \cite{hir87,bio87}.

Stars more massive than $\sim$20M$_\odot$ may have different fates.  
They have larger iron cores and will be intrinsically too massive to 
produce a neutron star via the supernova explosion. Then, the outcome will be a 
black hole. 
The gravitational collapse of these massive stars is currently attracting great interest. 
This is mainly because they are supposed to be associated with 
gamma ray bursts with long durations, one of the most energetic explosions in the universe.
Although the central engine and possible simultaneous production of
hyper-energetic Type Ic supernova (sometimes referred to as hypernova in the literature)
are remaining to be a mystery, many researchers believe that the collapse of 
{\it rapidly rotating} massive stars and the subsequent formation of black hole are responsible for
the phenomenon \cite{heg03}. 

Not all the massive stars may not be rotating so rapidly, though.
According to recent analysis of light curves of supernovae 
and nucleosynthetic yields by Nomoto and his collaborators \cite{mae03}, there appears to be a 
subset of supernovae with a progenitor mass of $\sim$20-25M$_\odot$, which produce a 
markedly smaller amount of $^{56}$Ni and, as a consequence, are substantially underluminous. 
They suggest that these faint supernovae are slow rotators. If this is the case, it is highly likely 
that there exist essentially {\it non-rotating} massive stars
that cannot produce even a dim optical display and form a black hole. They are the target of this Letter. 

Such a non-rotating, black-hole-forming collapse will be no less faint in neutrinos than the rotational 
counterpart. Following the core bounce and shock launch as in the ordinary supernova, there will be a 
phase of hyper-accretion through the stalled shock onto the proto-neutron star, lasting for $\sim$1 s 
before black hole formation. As shown below, as many neutrinos as in the ordinary supernova will be emitted 
during this period with distinctive characteristics in luminosities and spectra. Then the emission will be 
terminated rather abruptly after a black hole is formed. These features will be an unmistakable 
indication of black hole formation. The event will be unique in that they will be bright predominantly
by neutrinos. Hence they should be an important target of neutrino astronomy.

In fact, this channel of black hole formation (failed supernova), in particular, its
neutrino signal has not been studied in detail with hydrodynamical
simulations so far. 
Most of the previous papers were concerned with quasistatic evolutions of
mass-accreting proto-neutron stars \cite{bur88} or the so-called delayed collapse of proto-neutron stars \cite{gle95,bau96a,bau96b,pon99,pon01a,pon01b}
that is supposed to occur after a {\it successful} supernova explosion,
triggered by some phase transitions
%%  or 
%% a slow down of rapid rotation of proto neutron star
 during its cooling phase. (Note we do not consider  
supermassive or intermediate-mass black holes \cite{woo86} in this
Letter.) In some other papers \cite{bru01,lie04}, long term
hydrodynamical simulations of failed explosion similar to ours were
performed but their foci were different from ours. Hence this is the first serious and quantitative 
investigation into the phenomenon.

The most important findings in this Letter are the fact that the neutrino signals from the gravitational collapse of 
non-rotating massive stars are highly sensitive to the difference of equation of state (EOS) and can be used as a 
useful probe into the properties of hot and dense hadronic matter. 
EOS of dense matter is one of the most important ingredients in high energy astrophysics but is arguably
difficult to obtain from first-principle calculations and, as a result, the theoretical predictions are subject to 
uncertainty. Hence many attempts have been made over the years to obtain some 
constraints on EOS. We propose in this Letter to add a new item to this attempt-list, that is, the detection of 
neutrino signals from the hyper-accretion phase preceding the black hole formation in the non-rotating collapse of 
massive stars.

A number of such events is admittedly highly uncertain but might be 
a substantial fraction (20$\sim$40$\%$) of that of the ordinary supernovae, 
depending on the mass range for the event and the initial mass function of massive stars.  
Then this event will be as important a target for neutrino astronomy as the ordinary supernova is currently 
thought to be. It is, therefore, a very urgent task to provide theoretical predictions for luminosities and spectra
of neutrinos. 

% progenitor
% the accretion of material and the profile of pre-supernova star.  
-- {\it Models. }--
We adopt the presupernova model of 40M$_\odot$ by Woosley and 
Weaver \cite{woo95}.  
%This is the most massive model in the series of presupernova models 
%which include the {\it standard} 15M$_\odot$ model for supernova explosion.  
The model contains the iron core of 1.98M$_\odot$, which is 
larger than the ordinary size of $\sim$1.4M$_\odot$ and 
warrants the black hole formation without explosion as its fate.  
We use the profile of central part of this model up to 3.0M$_\odot$ 
in baryon mass coordinate to describe the accretion of envelope matter
for a long time.  

We follow the dynamical evolutions by a general relativistic $\nu$-radiation-hydrodynamical code that solves 
the Boltzmann equation for neutrinos ($\nu_e$, $\bar{\nu}_e$, $\nu_{\mu/\tau}$ and $\bar{\nu}_{\mu/\tau}$) 
together with lagrangian hydrodynamics under spherical symmetry \cite{yam97,yam99,sum05}.  
%The numerical code has been successfully used to study the long-term evolution 
%of supernova core in a 15M$_\odot$ star up to $\sim$1 s after 
%bounce \cite{sum05}. 
The general relativity is especially crucial in the current study, 
since the re-collapse is triggered by general relativity.  
We assume the spherical symmetry, targeting 
the massive-end objects of the faint supernova branch, which are supposed 
to be rotating slowly.  
% for unveiling the mechanism of hypernovae and/or collapsars.  

In order to assess the influence of EOS, we employ two sets of realistic EOS   
by Lattimer and Swesty (LS-EOS) \cite{lat91} and by Shen et al. (SH-EOS) \cite{she98a}.  
They have been used in most of recent studies of the ordinary supernovae \cite{sum05}  and 
are the current standard in the society. The subroutine of LS-EOS is based on 
the compressible liquid drop model for nuclei immersed in dripped nucleons. 
The data table of SH-EOS is relatively new and constructed in the relativistic nuclear many body 
frameworks \cite{bro90} with the nuclear interactions constrained by the properties of unstable nuclei \cite{sug94} that are available recently in radioactive nuclear beams facilities.
The two sets of EOS are, as a result, different from each other 
in stiffness at high densities and symmetry energy in neutron-rich 
environment \cite{sum04}. For example, the incompressibilities are 180 MeV and 281 MeV and 
the maximum masses of cold neutron star are 1.8M$_\odot$ and 2.2M$_\odot$ for LS-EOS and SH-EOS, 
respectively. 

It should be noted here that both sets of EOS take into account only the nucleonic degrees of 
freedom and ignore possible existence of other constituents such as hyperons, pion- and kaon-condensations
and deconfined quarks. We do not intend to justify the negligence (see e.g. \cite{gle95,bau96a,bau96b,pon99,pon01a,pon01b} for their importance in 
the proto-neutron star cooling). It is, however, mentioned that this 
is essentially the first attempt to explore the EOS-dependence of the neutrino signals from the non-rotating 
collapse of massive stars leading to the black hole formation. Considering also the inherent 
uncertainties in the theories of these constituents other than nucleons, we think it is reasonable to start with 
the most basic EOS with nucleons alone, which will also serve as a reference for the future work with other 
degrees of freedom taken into account. 

The weak interaction rates for neutrinos are implemented 
by following the formulation of Bruenn \cite{bru85}. 
In addition included are the plasmon process and the nucleon-nucleon 
bremsstrahlung process which is calculated consistently with the effective nucleon mass in
the EOS table \cite{sum05}.
The other recent developments of neutrino-matter interactions \cite{bur05} 
and electron-capture rates \cite{hix03} are not included in the current computations
to facilitate the comparison with the previous study on 15M$_\odot$ star \cite{sum05}. 
It has been demonstrated that the electron-captures are important to determine the 
mass of the bounce core and the improvement may give a difference of
$\sim 0.1$M$_{\odot}$ \cite{hix03}.  
The implementation of these reaction rates and more consistent treatments of reactions
with EOS will be an important issue for the future work. We think, however, the basic scenario 
from the core bounce and shock stall through the hyper-accretion onto the proto-neutron star and 
then to the final collapse to black hole would not be changed substantially.

\begin{figure}
\includegraphics[scale=0.43]{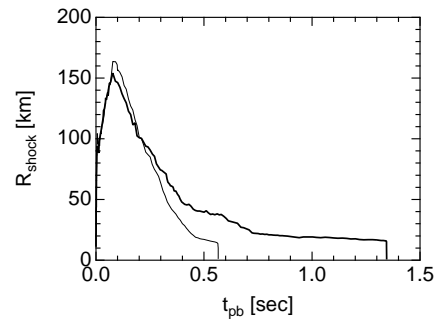}%
\caption{\label{shock} Position of shock wave as a function of time 
after bounce in models LS (thin) and SH (thick).}
\end{figure}
-- {\it Results. }--
We start the description of numerical results 
by presenting the radial position of shock 
wave as a function of time after the core bounce (t$_{pb}$) in Fig.~\ref{shock}.  
In both models, the launched shock wave reaches 
the maximum radius and then turns into recession around t$_{pb}\sim$100 ms 
owing to the ram pressure of falling matter.  
The size of bounced core is similar in two models 
and does not give much difference in the early dynamics of shock wave.  
Clear differences appear in the recession of shock wave and 
the shrinkage of central core after t$_{pb}$=100 ms when the hyper-accretion phase 
sets in. It should be noted that the accretion rate for the present model of 40M$_\odot$ 
($\sim$1 M$_\odot$/s at t$_{pb}$=0.4 s) is considerably higher 
than that ($\sim$0.2 M$_\odot$/s) for the canonical 15M$_\odot$ model for supernovae.
This fact results in much faster contraction of central cores in the former.  

In model LS, the shock wave recedes quickly down to $\sim$20 km.  
The central core contracts rapidly as its mass
increases toward the maximum value for hot and lepton-rich configurations in stable equilibrium.  
At t$_{pb}$=0.56 s, a dynamical collapse finally sets in and the central core shrinks on a dynamical
time scale. By this time, the enclosed baryon mass and gravitational mass inside the shock wave 
reach 2.10M$_\odot$ and 1.99M$_\odot$, respectively.  
Within the next $\sim$8 ms, the central core becomes compact enough to form
an apparent horizon at $\sim$5 km, which marks the formation of black hole.  
%The lapse function, $e^{\phi}$, and the general relativistic Lorentz 
%factor, $\Gamma$, 
%decrease to $\sim$0.4 at around the apparent horizon.

In model SH, on the other hand, the shock wave recedes 
rather slowly over $\sim$1 s. The dynamical collapse starts when 
the enclosed baryon mass reaches 2.66M$_\odot$ (gravitational mass 2.38M$_\odot$) 
at t$_{pb}$=1.34 s.
%The dynamical collapse is faster than LS and 
%The apparent horizon is formed at $\sim$5 km within $\sim$1 ms.  
% neutrino-sphere swallowed, signal fade out, Baumgarte
This remarkable difference in the durations of the hyper-accretion phase preceding the black hole 
formation is worth particular emphasis. It originates mainly from the difference in the maximum 
mass of the hot and lepton-rich core in stable equilibrium and, to lesser extent, from the difference
in the accretion rates. Hence, if observed, this will provide us with invaluable information on the stiffness
of EOS.
% maximum mass different from cold NS, thermal?

\begin{figure}
\includegraphics[scale=0.36]{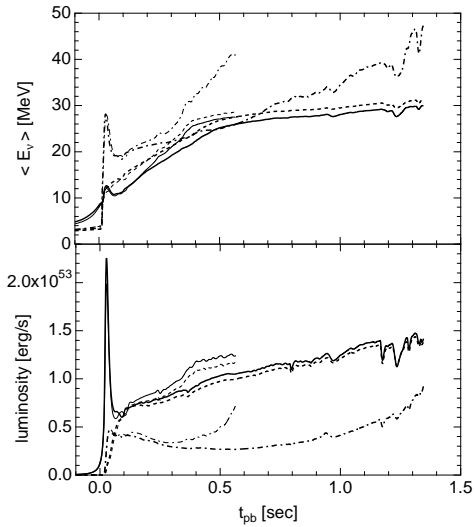}%
\caption{\label{neutrino}Average energies (upper) and luminosities (lower)
of $\nu_e$ (solid), $\bar{\nu}_e$ (dashed) and $\nu_{\mu/\tau}$ (dot-dashed) 
as a function of time (t$_{pb}$) in two models LS (thin) and SH (thick).}
\end{figure}
This novel difference is most clearly reflected in the duration of neutrino emissions   
as demonstrated in Fig.~\ref{neutrino}, where the average energies and luminosities of neutrinos are 
shown as a function of time (t$_{pb}$). The end points in the figure correspond to 
the formations of apparent horizon, i.e. the births of black hole. Note, however, that the major decline 
of neutrino emission will occur a fraction of millisecond later when the neutrino sphere is swallowed
by the horizon and will be recognized at the boundary ($\sim$6000 km) another $\sim$20 ms 
later, when neutrinos outside the neutrino sphere have traversed the distance at the light velocity. 
Unfortunately, we cannot follow this termination of neutrino emission owing to numerical problems.
We will have to implement a scheme to avoid both coordinate- and real singularities to handle this 
problem (see, e.g., \cite{bau96a,bau96b}). However, it is stressed that this problem does not matter in this 
Letter. The point here is that the longer-term neutrino emissions during the hyper-accretion phase is 
more revealing.

%The end point of neutrino emission at t$_{pb}$=0.57 s 
%in model LS corresponds to the formation of apparent 
%horizon, i.e. the birth of black hole.  
%In model SH, on the other hand, the neutrino emission 
%continues over 1 s and ends at 1.35 s.  
%Note that, observationally, neutrino emission continues further 
%for a while ($\sim$20 ms for a observer at $\sim$6000 km), 
%during which neutrinos outside the neutrino sphere travel, 
%and then the neutrino signal is terminated.  
%Therefore, the different timing of the black hole formation 
%corresponds to the different timing of the termination of neutrinos.  
%This fact will clearly provides us with the information on EOS.  
%
%In addition, the increase of average energies and 
%luminosities in time are faster in LS than those in SH, 
%though the qualitative behavior is similar.  

The time profile of luminosities 
right after bounce is similar to the ones in ordinary 
supernovae having the neutronization burst of $\nu_e$ 
and the rise of  $\bar{\nu}_e$, $\nu_{\mu/\tau}$ 
and $\bar{\nu}_{\mu/\tau}$.  
Luminosities afterward are dominated by the contributions 
from the accreted matter, which is heated up by the shock wave and 
further by compression onto the proto-neutron star surface.
Since the accreted matter contains a lot of electrons and positrons, they annihilate with each other
to create pairs of neutrino and anti-neutrino of all species. They are also captured by nucleons to 
produce electron-type neutrinos and anti-neutrinos. This latter processes are responsible for the 
dominance of $\nu_e$ and $\bar{\nu}_e$ as well as their similarity in the luminosity.

The difference in the reactions also leads to the difference in the radial positions of neutrino sphere 
and, hence, to the hierarchy of average energies shown in Fig.~\ref{neutrino}.  
The average energy of $\nu_{\mu/\tau}$ and $\bar{\nu}_{\mu/\tau}$ 
is particularly a good indicator of the difference of temperatures in two models, 
having, for example, a higher average energy at t$_{pb} \sim$0.5 s in model LS owing to the faster contraction.  
It is remarkable that the luminosities and average energies increase by a factor of two or more 
toward the formation of black hole, which will be utilized for diagnosis of the present 
channel of black hole formation.

\begin{figure}
\includegraphics[scale=0.46]{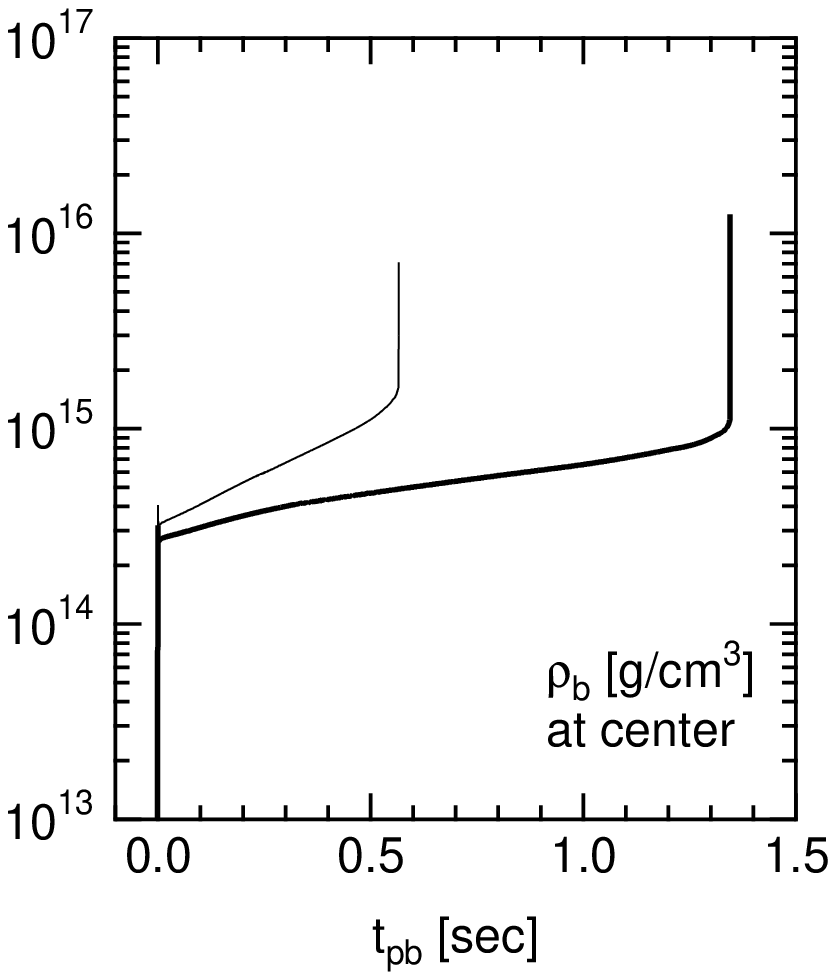}%
\hspace{-1.0cm}
\includegraphics[scale=0.46]{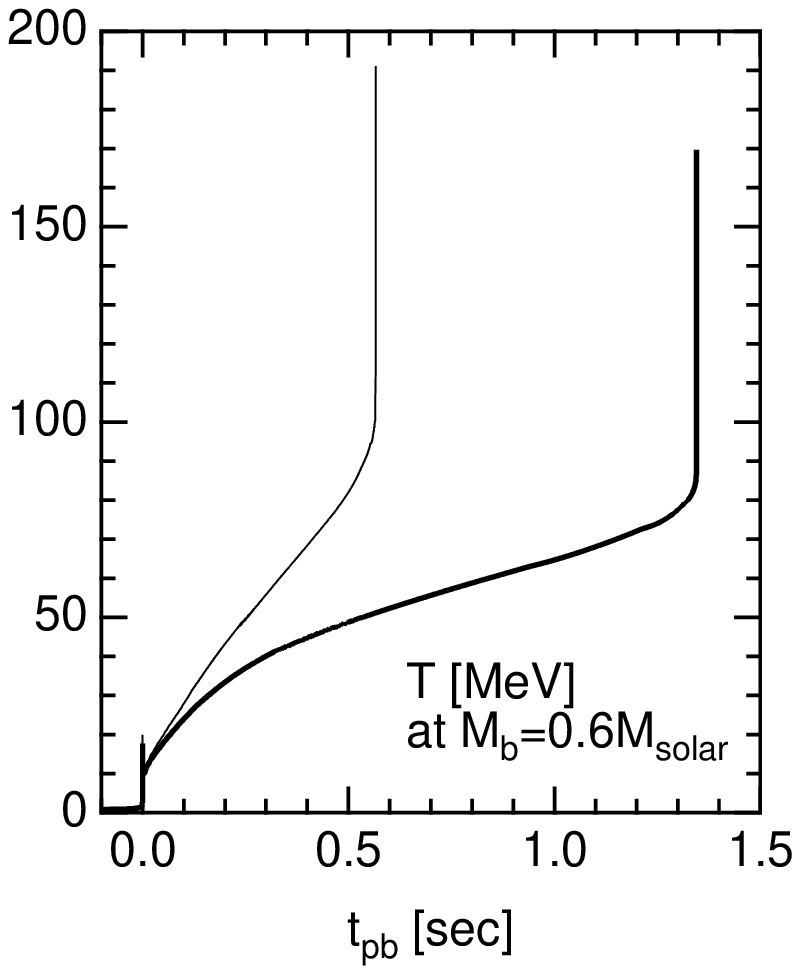}%
\caption{\label{density}Baryon mass density at center (left) 
and temperature at off-center (right) %at 0.6M$_\odot$ in baryon mass coordinate
as a function of time (t$_{pb}$) 
in models LS (thin) and SH (thick).}
\end{figure}
The earlier contraction of proto-neutron star and 
formation of black hole in LS apparently comes from the softness 
of the EOS, which gives a smaller maximum mass of neutron star.  
%This tendency was seen in the previous study of 15M$_\odot$ star \cite{sum05}.  
%In the current case of 40M$_\odot$ star, general relativity 
%enhances the tendency further.  % accretion rate
In Fig.~\ref{density}, we display the evolution of central density 
as a function of time (t$_{pb}$).  
Right after a small peak at bounce, 
the central density in model LS rises very quickly  
toward the final collapse in contrast 
to the much slower increase in model SH.
%% The central density in LS goes beyond 10$^{15}$ g/cm$^3$ 
%% ($\sim 3\rho_0$) at t$_{pb}$=0.46 s and reaches 
%% $\sim$7$\times$10$^{15}$ g/cm$^3$ at the end.
%%
Since the contraction of proto-neutron star proceeds almost adiabatically,
the temperature inside becomes enormously high owing to the compression.  
The peak temperature is attained off center near the inner core surface thanks 
to the shock heating. We show in Fig.~\ref{density} the time evolution of temperature 
at 0.6M$_\odot$ in baryon mass coordinate, which is near the temperature peak.
The temperature reaches around 100 MeV at the beginning 
of re-collapse and exceeds much over 100 MeV for both models while it
rises much slower before re-collapse in model SH.
%% In model LS, the temperature reaches 95 MeV at the beginning 
%% of re-collapse and ends up with 191 MeV while it reaches 170 MeV eventually 
%% after the much slower rise in model SH.
%and the peak of temperature profile 
%is 86 MeV at the beginning of re-collapse and ends up 
%at 170 MeV.  

It is to be noted that we have used LS-EOS and SH-EOS 
in very high density and temperature regimes that they may not have been 
originally meant for. (We have extended the SH-EOS table using the original framework 
for the current simulations.)  
This will certainly be an over-simplification.  Exotic phases such as meson condensations, hyperons or
deconfined quarks are expected to appear at certain densities and temperatures.  
The current models will serve as a reference for comparison. 
If, for example,  a new phase emerges at a certain point in evolution, 
the softening of EOS will cause immediately a dynamical collapse to black hole just like the
ordinary delayed collapse. Hence the present simulations provide upper limits to the duration of neutrino 
emission before black hole formation. It should be emphasized again that the difference in EOS is already
remarkable in the neutrino signals without these exotic constituents, which is in good contrast to the 
proto-neutron star cooling.
%To enable this EOS probe feasible, one has to predict neutrino 
%spectrum quantitatively using the established microphysics 
%as well as signal template for ordinary supernovae.  
%It is also essential to detect neutrino signals from 
%the beginning of neutronization burst together with the 
%rise of neutrino energies.  
%
%EOS at high densities and high temperatures
%hyperons, quark degrees of freedom.  
%Hadron physics.  

The current generation of neutrino detectors will detect 
thousands of neutrinos from the above-discussed events in our galaxy. 
Simultaneous observations of different neutrino flavors,  
from the initial burst of $\nu_e$ through the rise of luminosity and hardening of spectra 
up to the termination, at Super-Kamiokande, SNO
%%the Sudberry Neutrino Observatory
and 
other facilities together with an appropriate consideration of neutrino oscillation are indispensable to claim the black hole formation 
through the channel of current interest. 
The neutrino signal may be a complex folding of a couple of factors 
in addition to EOS.
It is of urgent importance to provide more detailed
theoretical predictions, particularly the dependence not only on EOS 
but also on the mass and structure of progenitor, which we are 
currently doing systematically.  
%To estimate a number of black-hole-forming supernovae more 
%precisely, we need systematic studies since the fraction 
%is sensitive to the mass distribution of progenitors.  
%Upper limit of the mass range for the black hole formation 
%by accretion is yet uncertain and depends on the metallicity \cite{nak05}.  
%It is also necessary to study the outcome for rapidly rotating massive 
%stars to consider the other types of black hole formation 
%and associated phenomena.  
%
% overflow test
%The neutrino signal may be a complex folding of factors such as 
%progenitor models in addition to EOS.
%The neutrino signal may be a complex folding of factors such as.

In summary, we computed neutrino signals from the collapse of a non-rotating 
massive star of 40M$_\odot$ and found that they are remarkably sensitive to the difference of
EOS and, hence, can be utilized as a novel probe into the properties of hot and dense matter. 
The terrestrial detection of such events in future will reveal the 
quantitative details of the accretion and contraction of proto-neutron star deep inside the star 
leading to black hole formation, and will give a new constraint on EOS at high density and temperature.  
%Nucleosynthesis by rotating case, fall back.

%acknowledgment
The numerical simulations were performed 
at NAO/ADAC (wks06a, wkn10b) 
and JAERI, and were done partially 
at KEK (No.\ 126) and RIKEN.  
This work is partially supported by the Grants-in-Aid for the 
Scientific Research (14039210, 14079202, 15540243, 15740160, 17540267)
of the MOESSC of Japan, 
%Ministry of Education, Science, Sports and Culture of Japan, 
for 
the 21st century COE program "Holistic Research and
Education Center for Physics of Self-organizing Systems", and for
Academic Frontier Project of MEXT.

\end{document}